
%
%
%
%
\hsize=13cm
\hfuzz=20pt
\vsize=18cm
\tolerance=10000
\magnification=1200
%

\font\headfont=cmbx10 scaled 1440
\font\namefont=cmr10
\font\initialfont=cmr10 scaled 1200
\font\addfont=cmti10
\font\fntefont= cmr7 scaled 1200
\def\fracfont#1{{\the\scriptfont0 #1}}


\def\sp{\ }
\def\seq{\sp=\sp}
\def\pls{\sp+\sp}
\def\mi{\sp-\sp}
\def\pd{\partial}
\def\mbox#1#2{\vcenter{\hrule width#1in\hbox{\vrule height#2in
   \hskip#1in\vrule height#2in}\hrule width#1in}}

\def\frac#1#2{{\fracfont{{#1}\over{#2}}}}
\def\frc#1#2{\leavevmode\kern .1em
             \raise .5ex\hbox{\the\scriptfont0 $#1$}\kern -.1em /
             \kern-.15em\lower .25ex\hbox{\the\scriptfont0 $#2$}}
\def\half{\frac{1}{2}}
\def\inv#1{\frac{1}{#1}}
\def\oplu#1{\leavevmode\raise .3em\hbox{$\oplus$}
            \kern-1.51em\lower.4em\hbox{\the\scriptfont0 #1}}
\def\nrmord#1{\leavevmode\raise .3em\hbox{\the\scriptfont0 $#1$}
            \kern-.43em\lower.3em\hbox{\the\scriptfont0 $#1$}}

\def\slash#1{\hbox{$#1$\kern-.52em\hbox{$/$}}}

%
%
\def\drawline#1{{\hbox to #1truein{\hrulefill}}}   

\def\fnote#1#2{\baselineskip 14pt{\footnote{$^#1$}{\fntefont #2}}
               \baselineskip 18pt}
\count101=0
\def\fnoter#1{\advance\count101 by 1
              \baselineskip 14pt{\footnote{$^{\backslash \number\count101
              }$}
                                            {\fntefont #1}}
               \baselineskip 18pt}
%
%
\def\title#1#2#3{\centerline{\bf{\headfont #1}}
			\medskip
			\centerline{\bf{\headfont #2}}
			\medskip
			\centerline{\bf{\headfont #3}}
		       }
\def\contract{$\!$
	\fnote{\star}{This work is supported in part by funds provided by the
		      U. S. Department of Energy (D.O.E.) under contract
		      \#DE-AC02-76ER03069.
		     }
	     }
\def\author#1#2#3#4#5{\vskip 0.5truein
	\centerline{{\initialfont #1}{\namefont #2}{\initialfont #3}
		    {\initialfont#4}{\namefont#5}}
		     }
\def\address{\medskip
	\centerline{\addfont{Center for Theoretical Physics,}}
	\centerline{\addfont{Laboratory for Nuclear Science}}
	\centerline{\addfont{and Department of Physics}}
	\centerline{\addfont{Massachusetts Institute of Technology}}
	\centerline{\addfont{Cambridge, Massachusetts 02139 U.S.A.}}
	    }
\def\abstract{\vskip 0.75truein
		\baselineskip 18pt plus 2pt minus 2pt
		\centerline{\bf ABSTRACT}
		\medskip
		\par
		}
\def\ctp_no#1#2{\noindent CTP \#{#1}\hfill #2}
%
%
\def\today{\ifcase\month\or
  January\or February\or March\or April\or May\or June\or
  July\or August\or September\or October\or November\or December\fi
  \space\number\day, \number\year}

\def\datetime{

        \count100=\time                               
        \count110=\count100

        \divide\count110 by 60                        
        \count200=\count110
        \multiply\count110 by 60

        \advance\count100 by -\count110               
        \count250=\count200                           
        \advance\count200 by -12                      



        \noindent {\bf PRELIMINARY VERSION}
        \hfill{\bf \today} \qquad\qquad                
        \ifnum \count200 < 0 {\bf \number\count250}    
                     \else {\bf \number\count200}\fi
        \ifnum \count100 < 10 {\bf :\sp0\number\count100}
                     \else {\bf :\sp\number\count100}\fi
        \ifnum \count200 < 0 \sp{\bf A.M.}                       
                     \else \sp{\bf P.M.}\fi

        }
%
%

%
%
\def\refs{\vfill\eject
	  \centerline{\bf REFERENCES}
	  \vskip 0.9truecm}
\def\ref#1#2{\item{#1.}{#2.}\smallskip\goodbreak}

\def\np#1{{\it Nucl. Phys.} {\bf B#1}}
\def\cmp#1{{\it Commun. Math. Phys.} {\bf #1}}

%

\def\b{\beta}

\def\d{\delta}

\def\g{\gamma}

\def\k{\kappa}
\def\l{\lambda}

%
%
\def\ca{{\cal A}}

\baselineskip 15pt

\def\pd{\partial}
\def\pdb{{\bar\partial}}
\def\zbar{{\bar z}}
\def\frac#1#2{{{#1}\over{#2}}}
\def\inv#1{{1\over{#1}}}
\def\ev#1{\langle #1\rangle}
\def\op#1{e^{i #1}}

\def\ca{{\cal A}}

\def\varphiI{\varphi_{(I)}}
\def\d{\delta}

\def\ddz#1{\d^{(2)}(#1)}
\def\us{u^*}
\def\metric{g_{z\zbar}}
\def\kt{\tilde\k}

\def\REFewTSM{1}
\def\REFmsTS{2}
\def\REFvvTG{3}

\def\EQABS{1}
\def\EQA{2}
\def\EQAP{3}
\def\EQSTS{4}
\def\EQCHOOSE{5}
\def\EQCHI{6}
\def\EQZTW{7}
\def\EQZEROM{8}
\def\EQGZZ{9}

\def\EQZTS{11}
\def\EQSTSB{12}
\def\EQCTS{13}
\def\EQRES{14}
\def\EQVGEN{15}
\def\EQRESGEN{16}

\title{Relating Scattering Amplitudes in}
	{Bosonic and Topological String Theories\contract}
	{}

\author{R}{OGER}{}{B}{ROOKS}

\address

\abstract

A formal relationship between scattering amplitudes in critical bosonic
string theory and correlation functions of operators in topological string
theory is found.
\vskip 7truecm
\ctp_no{2080}{March 1992}

\vfill\eject

At its essence, string theory is a locally reparametrization
invariant\fnoter{In this paper we will only be working with the critical
bosonic string theory.} theory in $1+1$ dimensions.  One of its interesting
features is that correlation functions on the world-sheet yield scattering
amplitudes for states in the space-time theory.  In this vein, we might ask if
the bosonic  string is the unique world-sheet theory which computes space-time
scattering amplitudes.  That is, might there be another (and perhaps
technically simpler) reparametrization invariant theory whose correlation
functions might be mathematically interpreted as the scattering amplitudes for
states we identify as belonging to a string theory.

One of the crucial features of the bosonic string is a direct consequence of
its world--sheet reparametrization invariance.  As the metric only has three
components, the three local parameters of the reparametrization symmetry makes
the metric a pure gauge degree of freedom.  After local gauge fixing, the
moduli of the Riemann surface remain.  It so happens that this is all a
built--in feature of topological quantum field theories coupled to
two-dimensional topological gravity.

Topological gravity theories can be constructed as a gauge fixing of the local
symmetry $\d g_{ab}={\hat\Psi}_{ab}$ on the metric $g_{ab}$. This is a
symmetry of the Einstein--Hilbert action in two--dimensions as the latter
happens to be the density for the Euler characteristic. Thus the metric is a
pure gauge degree of freedom modulo boundary terms. However, for gauge fixing
conditions with zero-mode solutions, interesting interpretations can be made
of the resulting construction.

Of principal importance in string theory are the maps from the Riemann surface
to the target or space-time manifold; the coordinates of the string. The
scattering amplitudes for the string states are given by correlation functions
of two dimensional vertex operators built out of these space-time coordinates.
In particular, the vertex operators are written as integrals over the Riemann
surface of operators with total conformal dimension equal to two. Thus, the
positions at which the states puncture the Riemann surface are integrated out
in the scattering amplitudes.  Formally, the simplest operator is the one
which creates tachyons.  It is written as $V_T[k,x,g]=\int_\Sigma \op{k\cdot
x}$ where $k_\mu$ is the momentum of the tachyon and $\metric$ is the fiducial
metric on the Riemann surface, $\Sigma$.   Should the scattering amplitudes
for the bosonic string be computable in another theory, the identification of
the vertex operators (in that theory) should be unambiguous.

As mentioned earlier, topological theories are independent of the local
degrees of freedom of the metric.  As in string theory we can introduce fields
which act as maps from the Riemann surface to the target manifold, $M_{TS}$.
Unlike the bosonic string, this coupled theory is manifestly constructed to be
topological with respect to the Riemann surface.  That is, the only physical
states in the theory correspond to operators whose correlation functions yield
cohomological information about the moduli space of the Riemann surface and/or
the moduli space of particular maps (holomorphic, for example) from $\Sigma$
to $M_{TS}$.  The example of such a theory which looks most like a string
theory is known in the literature as the topological string theory
[\REFewTSM,\REFmsTS]. It is obtained by either ``twisting'' a locally $N=2$
supersymmetric string theory or via a BRST gauge fixing of local topological
symmetries on the coordinates $u^A$ ($\d u^A={\hat \chi}^A$), which map
$\Sigma$ to $M_{TS}$, and the metric on the Riemann surface ($\d
g_{ab}={\hat\Psi}_{ab}$).  In the former case, $M_{TS}$, must be K{\" a}hler.
However, in the BRST approach this condition may be relaxed and $M_{TS}$ may
be taken to be an arbitrary almost complex manifold. In this paper, we will
take $M_{TS}$ to be flat with real dimension $D$.  We will choose a value for
$D$ later\fnoter{There is no anomaly in the topological string with flat
target space.}. The ``twisting'' amounts to a declaration that the fermionic
fields $\bar\psi$ and $\psi$ of the $N=2$ superstring are now dimension $2$
and $0$ fermionic fields respectively: $(\bar\psi,\psi)\to(\rho,\xi)$.
Similarly, the Grassmann even superghosts $(\beta,\g)$ will be of dimension
$(2,-1)$ (we will not change their names here).

As it is a topological theory, the correlation functions of observables
(members of the super (BRST) charge cohomology) in this theory are independent
of local world-sheet data.  Thus it is suggestive that there might be some
relation between the scattering amplitudes in critical bosonic string theory
and the correlation functions of topological string theory. However, we are
quickly sobered by the fact that the states of the bosonic string are not
members of a de Rham or Dolbeault cohomology in space-time. So it would seem
that the best we can hope for is that the correlation functions of some more
general operators are mathematically the same as scattering amplitudes in the
bosonic string.  Then, exploiting the large number of symmetries of the
topological string, we might be able to relate these correlation functions to
topological invariants via Ward Identities.  The first step of this algorithm
will be the topic of this paper. We will concentrate on tachyon scattering
amplitudes in the critical bosonic string as the arena for the development of
our program. We will then see how straightforward it is to generalize our
result to the scatterings of other states. Our work will be performed on the
sphere.  However, we expect that our results may be generalized to higher
genus surfaces.

The path integral expression for the scattering amplitude of $N$ tachyons is
$$\ca(k)\seq {\ev{\prod_{I=1}^N V_{T}[k_I,x,g]}_{BS}}\seq \int
[dg][dx]\prod_{I=1}^N \int d^2z_I\sqrt{g(z_I)} \op{k_I\cdot
x(z_I)}e^{-S_{BS}}\ \ .\eqno(\EQABS)$$
Here, $S_{BS}$ is the action for the critical bosonic string:
$S_{BS}=\half\int d^2 z\sqrt{g}g^{ab}\pd_a x\cdot \pd_b x$. As mentioned
earlier, we have enough gauge freedom to locally fix the metric completely.
It is convenient to work on the complex plane.  Our amplitude then reads
$$\ca(k)\seq \int[dx]\prod_{I=1}^N \int d^2 z_I \op{k_I\cdot
x(z_I)} e^{-S_{BS}}\ \ ,\eqno(\EQA)$$
with $S_{BS}[x]=\half\int d^2z \pd x\cdot \pdb x$.  It is customary to
re-write this expression as
$$\eqalign{
\ca(k)\seq&\prod_{I=1}^N \int d^2 z_I
\int [dx] e^{-S'_{BS}[k,x]}\ \ ,\cr
S'_{BS}\seq& \half\int d^2 z \partial x\cdot \pdb x \mi i\int d^2
z \sum_{I=1}^N k_I\cdot x(z)\ddz{z-z_I}\ \ .\cr}\eqno(\EQAP)$$

We seek an expression which is formally the same as $\ca(k)$ in the
topological string theory.  We will not fully describe the manner by which the
action for the topological string is derived as this is adequately discussed
in the literature [\REFewTSM,\REFmsTS].  Suffice it to say that at an
intermediate stage one arrives at the following (super) scale invariant
action, in the super-conformal gauge.
$$
S_{TS}\seq \int_\Sigma (\half \pd u\cdot\pdb \us \pls \rho\cdot\pdb\xi \pls
b\pdb c \pls \beta\pdb\g \pls {\rm c.c})\ \ .\eqno(\EQSTS)$$
This is the twisted version of the $N=2$ superstring but in arbitrary
space--time dimensions.  The field content is as follows. The $D$ sets of
fields $\lbrace u,\xi,\rho\rbrace$ form topological (super) multiplets; so do
the fields $(b,\b)$ and $(c,\g)$.  The dimensions of the bosonic fields are
$[u]=0$, $[\b]=2$, $[\g]=-1$ and those of the fermionic fields are $[\xi]=0$,
$[\rho]=1$, $[b]=2$ and $[c]=-1$.

Let us now focus our attention on the scale invariance in our action. With
$\varphi\equiv \ln\metric$, the scale transformation corresponds to an
arbitrary local shift of $\varphi$.  As this is a purely algebraic
transformation (no derivatives involved), $\varphi$ may be chosen to be of
whatever form is convenient to the problem at hand. In string theory it is
customary to simply set $\varphi=0$.  The counterpart of this transformation
in the topological string theory reads $\d\varphi={\hat \psi}$.  In the $N=2$
supersymmetry language $\hat \psi$ becomes the superpartner of $\varphi$.  In
the BRST language, $\hat\psi$ will become a scalar but fermionic ghost field.

Following the BRST approach, impose that the curvature of the super-conformal
gauge metric is equal to a fixed curvature, $\hat R$.  Our only condition on
$\hat R$ is that it be of the topological class of the Riemann surface the
theory is defined on.  So we take $\inv{2\pi}\int_\Sigma {\hat R} =\chi$ to be
the Euler characteristic of $\Sigma$. In order to relate $\ca(k)$ in equation
(\EQAP) to a correlation function in the topological string, we will find it
useful to choose the gauge [\REFvvTG]
$$\pd\pdb \ln{g_{z\zbar}}\seq -g_{z\zbar} R\sp\equiv\sp \sum_{I=1}^N \kappa_I
\ddz{z - z_I}\ \ .\eqno(\EQCHOOSE)$$
It then follows that
$$\sum_{I=1}^N \kappa_I\seq 2\pi\chi\ \ .\eqno(\EQCHI)$$
Then the partition function for the gauge fixing of the scale invariance reads
$$\eqalign{
Z_{TW}\seq& \int [d\metric] [d\l][d\eta][d\psi] e^{-S_{TW}}\ \ ,\cr
S_{TW}\seq& \int d^2z \l[\pdb\pd \varphi \mi \sum_{I=1}^N\kappa_I\ddz{z-z_I}]
\pls \int d^2z \eta\pdb\pd\psi\ \ .\cr}\eqno(\EQZTW)$$
In the action $S_{TW}$, the first term is the gauge fixing term and the second
is the corresponding ghost action.  Note that it is quadratic in derivatives.
The field $\l$ is a Lagrange multiplier (BRST auxiliary field) which imposes
the constraint (\EQCHOOSE) while $\eta$ and $\psi$ are dimensionless fermionic
fields. These fields form the following topological multiplets:
$(\varphi,\psi)$ and $(\eta,\l)$.  To compute this partition function we must
first remove $N$ $\varphi$ zero-modes which we label by $\varphiI$.  Their
equations are
$$\pdb\pd\varphiI\seq \kappa_I \ddz{z-z_I}\ \ .\eqno(\EQZEROM)$$
We know this as the equation for the scalar Green's function with solution
$$\varphiI\seq \kt_I\ln{|z-z_I|}\qquad \Longrightarrow\qquad \metric^{(I)} \seq
|z-z_I|^{\kt_I}\ \ ,\eqno(\EQGZZ)$$
where $\kt_I=\frac{\k_I}{2\pi}$.
Having removed these zero-modes, the integrals over the Lagrange multiplier,
the $\varphi$ non-zero-mode and the ghost fields may be done.  The resulting
determinants cancel each other. As the $\varphi$ zero-modes are in one-to-one
correspondence with the locations of the singularities we then replace
$\int[d\metric]$ by $\prod_{I=1}^N\int d^2z_I$ up to a Jacobian factor.
Altogether, the full partition function of the topological string now reads
$$Z_{TS}\seq \prod_{I=1}^N  \int d^2z_I J(\k_I,z_I)\int
[du][d\rho][d\xi][db][dc][d\b][d\g] e^{-S_{TS}}\ \ ,\eqno(\EQZTS)$$
where $J$ is the Jacobian for $\metric^{(I)}\to z_I$. It is clear that the
integrals over the locations of the singularities are needed as these points
are, after all, gauge artifacts.

Now, when we bosonize the fields which appear in first order form in $S_{TS}$
we find
$$S_{TS}\seq \int_\Sigma (\half \pd u\cdot\pdb\us \pls \half
\pd\phi\cdot\pdb\phi \pls i\inv{8} \sum_{I=1}^N \kappa_I\ddz{z - z_I}
q\cdot\phi(z))\ \ .\eqno(\EQSTSB)$$
Here we have replaced the set of $(\rho,\xi)$ fields by a set of $D$ real
bosons $\phi^i$, the $(b,c)$ system by the single real boson $\phi_{D+1}$ and
$(\b,\g)$ by $\phi_{D+2}$.  Due to the fact that the $(\beta,\gamma)$ system
is bosonic, the metric on the space $(\phi^i,\phi_{D+1},\phi_{D+2})$ is ${\rm
diag}(+,\ldots,+,-)$ with the minus corresponding to the $\phi_{D+2}$
direction.  Whereas the bosonization of the fermionic partners of the target
manifold's coordinates do not change the Euclidean signature, the superghosts
give the new target manifold a Minkowskian signature with $\phi_{D+2}$ being
the time direction. Note that the dimension of this new space is $D+2$. The
background charges are $(q_i,q_{D+1},q_{D+2}) =(-1,-3,3)$.  The condition
(\EQCHOOSE) has been imposed.

Call the bosonic fields $\Phi\equiv\lbrace u,u^*,\phi\rbrace$ collectively.
The $\Phi$'s are real bosonic fields which we now interpret as the coordinates
of a Minkowski target manifold.  The dimension of this manifold is $3D+2$.  In
order the relate our correlation functions we must set this number equal to
the space-time dimension of the bosonic string: $3D+2=26$ or $D=8$

Consider computing the following correlation function in the topological
string
$$\eqalign{
\ev{\prod_{I=1}^N J^{-1}(\k_I,z_I)
e^{ip_I\cdot \Phi(z_I)}}_{TS} \seq& \prod_{I=1}^N \int d^2z_I
\int [d\Phi] e^{ip_I\cdot \Phi(z_I)} e^{-S_{TS}}\cr
\seq& \prod_{I=1}^N \int d^2z_I \int [d\Phi] e^{-S'_{TS}[p,\Phi]}\ \ ,\cr
S_{TS}'\seq \half\int d^2z \pd\Phi\cdot\pdb\Phi \mi& i\int d^2z \sum_{I=1}^N
p_I'\cdot \Phi(z)\ddz{z-z_I}\ \ ,\cr}\eqno(\EQCTS)$$
where $p_{I\mu}' \seq p_{I\mu}-\inv{8}\k_Iq_\mu$ and $q_\mu$ is zero in the
$u$ directions, $-1$ in the $\phi^i$ (or ($\rho,\xi$)) direction, $-3$ in the
$\phi_{D+1}$ (or $bc$) direction and $3$ in the $\phi_{D+2}$ (or ($\b,\g$))
direction.  We can now formally equate
$$\ev{\prod_{I=1}^N\int d^2z_I e^{ik\cdot x(z_I)}}_{BS}\seq \ev{\prod_{I=1}^N
J^{-1}(\k_I,z_I)e^{ip_I\cdot\Phi(z_I)}}_{TS}\ \ ,\eqno(\EQRES)$$
by declaring $p_{I\mu}=k_{I\mu}+\inv{8}\k_{I\mu}q_\mu$. This result may be
generalized to any bosonic string vertex operator of the form
$$V[k,\zeta,x]\seq \int d^2z P_\zeta(\pd x,\pd^2 x,\ldots)e^{ikx}\ \
,\eqno(\EQVGEN)$$
where $P_\zeta$ is a polynomial in powers of derivatives of $x$ and $\zeta$ is
a polarization tensor.  Then we obtain
$$\ev{\prod_{I=1}^N\int d^2z_I
P_{\zeta_I}(\pd x,\pd^2 x,\ldots)e^{ik\cdot x(z_I)}}_{BS}\seq
\ev{\prod_{I=1}^N J^{-1}(\k_I,z_I)
P_{\zeta_I}(\pd \Phi,\pd^2 \Phi,\ldots)e^{ip_I\cdot\Phi(z_I)}}_{TS}
\ \ ,\eqno(\EQRESGEN)$$
with $p_{I\mu}=k_{I\mu}+\inv{8}\k_Iq_\mu$.

Now that we have stated the formal relationship between the two theories, it
is left to the future to carry out the computations of the right hand side of
eqn. (\EQRESGEN) in the topological string theory.  The first step would be to
exploit Ward Identities or a Hodge decomposition of the inserted operators in
the topological string to relate the correlation functions to cohomology
classes.

We summarize by painting the following heuristic picture of what we have done.
We have given the Riemann surface of the topological string a curvature which
vanishes everywhere except for $N$ points, $z_I$, where it is singular. Then
bosonizing the fermions we have found that the background charge term mimicked
the source term in the bosonic string's scattering amplitudes.  The remaining
hurdle of taking the target manifold with Euclidean signature and making it
Minkowskian was scaled by bosonizing the topological gravity ghost systems.
The $\b$-$\g$ system was bosonic and so in bosonized form, its propagator
appeared with a sign opposite to the rest of the bosons.  Finally, operators
analogous to the vertex operators of the bosonic string  were introduced. An
operator was inserted at each of these points.  By construction, the manifold
had a curvature singularity wherever there is an operator insertion.  As these
points were gauge artifacts, we have found that we must integrate over them.
This was obtained by exchanging the integral over metric zero-modes for these
integrals as the zero-modes were in one to one correspondence with the $z_I$
points. The price of this is a Jacobian factor which may be absorbed into the
definition of the operators in the topological string.
\refs
\ref{\REFewTSM}{E. Witten, \cmp{118} (1988) 411}
\ref{\REFmsTS}{D. Montano and J. Sonnenschein, \np{313} (1989) 258}
\ref{\REFvvTG}{E. Verlinde and H. Verlinde, \np{348} (1991) 457}
\bye